\documentclass[aps,prl,reprint,twocolumn,preprintnumbers,floatfix,nofootinbib]{revtex4-1}

\usepackage[T1]{fontenc} 
\usepackage[utf8]{inputenc}
\usepackage[dvipdfm]{graphicx}
\usepackage{epstopdf}
\usepackage{mathrsfs}
\usepackage{amssymb}
\usepackage{verbatim}
\usepackage{color}
\usepackage[dvipsnames]{xcolor}
\usepackage{multirow}
\usepackage{amsmath}
\usepackage{subfig}
 \usepackage{slashed} 
\usepackage{float}
\usepackage[normalem]{ulem}
\usepackage{sidecap}
\usepackage{hyperref}
\usepackage{cancel}
\usepackage{enumerate}
\usepackage{tikz}
\usepackage[compat=1.1.0]{tikz-feynman}
\hypersetup{pdfstartview=FitV,colorlinks=true,linkcolor=blue,citecolor=red,filecolor=black,urlcolor=blue}

\newcommand*\diff{\mathop{}\!\mathrm{d}}

\newcommand{\beq}{\begin{equation}}
\newcommand{\eeq}{\end{equation}}
\newcommand{\bea}{\begin{eqnarray}}
\newcommand{\eea}{\end{eqnarray}}

\newcommand{\mrm}{\mathrm}
\newcommand{\trh}{T_{\mathrm{RH}}}
\newcommand{\tmax}{T_{\mathrm{max}}}

\def\to{\rightarrow}

\def\del{\partial}
\def\lrD{\overset{\leftrightarrow}{D}}
\def\lD{\overset{\leftarrow}{D}}
\def\rD{\overset{\rightarrow}{D}}

\usepackage[parfill]{parskip}
\begin{document}
\sloppy

\preprint{IFT-UAM/CSIC-21-13}
\preprint{KIAS-P21004}

\vspace*{1mm}

\title{Metastable Conformal Dark Matter} 

\author{Philippe Brax$^{a}$}
\email{philippe.brax@ipht.fr}
\author{Kunio Kaneta$^{b}$}
\email{kkaneta@kias.re.kr}
\author{Yann Mambrini$^{c}$}
\email{yann.mambrini@ijclab.in2p3.fr}
\author{Mathias Pierre$^{d,e}$}
\email{mathias.pierre@uam.es}
\vspace{0.5cm}

\affiliation{$^a$Institut  de  Physique th\'eorique  , Universit\'e  Paris-Saclay,CEA,  CNRS,  F-91191  Gif-sur-Yvette  Cedex,  France}

\affiliation{$^b$School of Physics, Korea Institute for Advanced Study, Seoul 02455, Korea}

\affiliation{$^c$Universit\'e Paris-Saclay, CNRS/IN2P3, IJCLab, 91405 Orsay, France}

 \affiliation{$^d$Instituto de F\'{i}sica Te\'{o}rica (IFT) UAM-CSIC, Campus de Cantoblanco, 28049 Madrid, Spain} 
\affiliation{
$^e$
Departamento de F\'{i}sica Te\'{o}rica, Universidad Aut\'{o}noma de Madrid (UAM), Campus de Cantoblanco, 28049 Madrid, Spain}

\date{\today}

\begin{abstract} 
We show that a metastable dark matter candidate arises naturally from the conformal transformation between the Einstein metric, where gravitons are normalised states,  and  the Jordan metric dictating the coupling between gravity and matter. Despite being secluded from the Standard Model by a large scale above which  the Jordan metric shows modifications to the  Einstein frame metric,  dark matter couples to the energy momentum tensor of
the Higgs field  in the primordial plasma primarily. This allows for the production
of dark matter in a sufficient amount which complies with observations. 
The  seclusion of dark matter  makes it  long-lived 
for masses $\lesssim 1$ MeV, with a lifetime much above
the age of the Universe and the present experimental limits. Such a dark matter scenario  has clear monochromatic signatures
generated by the  decay of the dark matter candidate into neutrino and/or $\gamma-$rays.

\vskip 1cm

\end{abstract}

\maketitle

\setcounter{equation}{0}

\section{I. Introduction}

Dark matter (DM) has now been  a mystery for more than 80 years. Ever since Zwicky's observation of the Coma Cluster \cite{Zwicky:1933gu}, the  measurements of Andromeda's rotation curve by Babcock \cite{babcock} and the issue of stabilising structures addressed by Peebles and Ostriker \cite{Ostriker:1973uit}, dark matter was systematically referred to as the  "subliminal matter problem" until Gunn {\it et al.} proposed in 1978 that the introduction of a new particle could fill the matter content of the Universe \cite{Gunn:1978gr}. Even if the "reality" of DM is now confirmed by the latest measurements of the CMB anisotropies \cite{planck},
it has taken a  long time to convince theorists and observers that the existence
a new field, beyond the Standard Model of particle physics, should exist
in order to explain the cosmological observations. This hypothesis, i.e. the presence of  a particle  in thermal equilibrium with the primordial plasma after the reheating phase,  has now become the most natural option for a large part of the physics community. Paradoxically, the contrary assumption that a highly feebly interacting candidate, the gravitino, could play the role of DM was one of the very first well-motivated candidate, proposed in \cite{pagels}, for the dark matter particle. The possibility that gravitino  could have been  in thermal equilibrium was contradicted in  \cite{gravitino} by taking into account its Planck reduced coupling to the thermal bath. Despite this early failure, a plethora of models based on the thermal equilibrium assumption, called WIMP for Weakly Interacting Massive Particle, were subsequently proposed (see \cite{Arcadi:2017kky} for a recent review on the subject). From Higgs-portal \cite{Higgsportal} to $Z$-portal \cite{Zportal}
 and $Z'$-portal \cite{Zpportal}, all the models based on this WIMP paradigm, which has the advantage of not questioning the earliest thermal stages of the Universe, are now becoming more and more in tension with the exclusion limits of the more recent direct detection experiments like XENON1T \cite{XENON}, LUX/LZ \cite{LUX} and PANDAX \cite{PANDAX}.

An alternative called FIMP for Feebly Interacting Massive Particle (or Freeze-In Massive Particle) was proposed in \cite{fimp}, where the dark matter component never becomes in equilibrium with the primordial plasma, and whose production rate is frozen "in" the process of reaching equilibrium. The original article deals with effective couplings  and can be seen as a generalisation of the gravitino dark matter,  which is of the same nature. This DM production mode has been since extended to High Scale SUSY models \cite{highscalesusy}, SO(10) constructions \cite{fimpso10}, $Z'$ mediators \cite{fimpzp}, heavy spin-2 and Kaluza Klein modes \cite{fimpspin2,fimpkk}, highly decoupled sectors \cite{fimphighlydecoupled,gravity}, or even in emergent gravity/string scenarios \cite{fimpemergent,fimpmoduli} (for a recent review, see \cite{Bernal:2017kxu}). All these proposals have in common the presence of higher dimensional operators at energies below a UV-scale $\Lambda$ determined by the  mass of the mediator or its  couplings (or both). As a rule, the presence of a cut-off scale at  higher values than the maximal temperature reached by the primordial plasma $\tmax$ (or the reheating $\trh$ if one considers instantaneous reheating) is conducive to  the FIMP mechanism in a dark matter sector. 

It is remarkable that such higher dimensional operators arise naturally in extensions of gravity, as it is the case in supergravity for instance. Originally, supersymmetry appeared as  an extension of the Poincaré group to spinorial transformations whose breaking generated  the neutrino as a goldstone fermion of the supersymmetry breaking \cite{va}. When supersymmetry becomes a local symmetry, i.e. supergravity, and after spontaneous supersymmetry breaking 
the longitudinal mode of the gravitino, also called the goldstino ($\Psi_{3/2}$), can be  considered as a dark matter candidate.
Its coupling to the 
Standard Model is obtained via its contribution to the metric by first defining an invariant vierbein under the generalized Poincaré transformations~\cite{va}
\begin{equation}
e_\mu^\alpha \ = \ \delta_\mu^\alpha -\frac{i}{2 F^2}
\left( \partial_\mu \bar \Psi_{3/2} \gamma^\alpha { \Psi_{3/2}} +  \bar \Psi_{3/2} \gamma^\alpha \partial_\mu {\Psi_{3/2}} \right)
\ , 
\label{va1}
\end{equation}
\noindent
$\sqrt{F}$ being related to the SUSY breaking scale\footnote{In this case, we can identify the cut-off scale of the model $\Lambda$
as $\sqrt{F}=\sqrt{m_{3/2} M_P}$, $m_{3/2}$ being the gravitino mass.}. In the absence of  an $R$-parity, $\Psi_{3/2}$ is a metastable neutral candidate whose spin-3/2
determines the final state ($\gamma+\nu$) of its decay products. 
This kind of construction belongs clearly to the category of models where the Standard 
Model fields interact with a dark sector through the presence of the {\it physical} or Jordan  metric $g^{\mu \nu} = e^\mu_a e^\nu_b \eta^{ab}$. Moreover, the suppression by $M_P$ of the extension of the metric makes $\Psi_{3/2}$ a perfectly long-lived FIMP candidate as argued above.

The idea of modifying the metric, or more precisely of considering that the geometrical metric $g_{\mu \nu}$, governing the gravitational structure and the propagation of gravitons differs from the metric governing the dynamics of  matter
$\tilde g_{\mu \nu}$, is not new. This was already proposed
 in Nordstrom gravitational theories~\cite{Nordstrom}, Brans-Dicke~\cite{Brans}
or Dirac's~\cite{Dirac}. Later a generalization 
to conformal and disformal transformations of the metric was introduced ~\cite{Bekenstein:1992pj,Bek2} and such
bimetric models became ubiquitous.
Coupling a scalar dark matter field via a conformal 
transformation of the metric
\beq
\tilde g_{\mu \nu} = e^{\phi / M_P}g_{\mu \nu}\simeq g_{\mu \nu}\left(1+\frac{\phi}{M_P}\right)\,,
\eeq
generating a coupling of the kind
\begin{equation}
    {\cal L}_{\phi}^\text{SM}\sim \frac{\phi}{2M_P}g_{\mu \nu}T^{\mu \nu}_\text{SM}\,,
    \label{eq:DMcouplingTmunu}
\end{equation}

\noindent
where $T^{\mu \nu}_\text{SM}$ represents the stress-energy 
tensor of the Standard Model,
may seem {\it a priori} dangerous as this induces dark matter's instability \cite{Choi:2019osi}. However, it is clear that the decay process are highly suppressed for $m_\phi \lesssim 1$ MeV as the only kinematically allowed final states are $\phi \rightarrow \nu \nu$ and loop-suppressed $\phi \rightarrow \gamma \gamma$, giving $\tau_{\phi\rightarrow \nu\nu}\sim \frac{M_P^2}{m_\nu^2m_\phi}\gtrsim 10^{36}$ seconds for $m_\phi \lesssim$ 1 MeV and $m_\nu\lesssim 0.05$ eV \cite{Dudas:2020sbq}. This
property is tightly related to the fact that the fermionic stress-energy tensor for an on-shell fermion $\nu$ is proportional to $m_\nu$. As result, one may ask oneself how to produce such a light dark matter candidate, with such a suppressed coupling to the Standard Model, in a sufficiently large amount to fulfill the cosmological abundance constraint. We will show that this is possible from scatterings involving the Higgs degrees of freedom, whose trace of the corresponding stress-energy tensor 
is proportional to the $\mrm{(energy)^2}$ stored in the plasma
which in turn can be very high at the end of the inflationary phase\footnote{More precisely of the order of $\sqrt{\rho_e}$, 
$\rho_e$ being the density of the inflaton at the end of inflation \cite{GKMO1,GKMO2}}, compensating the weakness of the Planck suppressed coupling. 
We will also study the dark matter produced via the decay of the inflaton since it was shown in \cite{Kaneta:2019zgw,inflatondecay} that this could dominate the production processes.
We will also explore the possibility to produce $\phi$ directly via inflaton decay.

{ Finally, notice that the effective coupling in Eq.~(\ref{eq:DMcouplingTmunu}) of a scalar particle to the trace of the energy momentum and suppressed by a large scale is analogous to constructions where scale invariance is broken spontaneously in a conformal sector coupled to a sector featuring explicit breaking terms~\cite{Bellazzini:2012vz}. In this case at low energy, the suppression scale can be identified as the typical vacuum-expectation-value (vev) breaking this
symmetry and the scalar particle is the associated pseudo Nambu-Goldstone boson.}

The paper is organized as follows. After a brief presentation of the model in section II, we will compute the relic abundance density of $\phi$, and its decay modes in section III. Section IV will be devoted to the analysis of the parameter space and smoking-gun signatures of our model before concluding in section V.

Throughout this
work, we use a natural system of units in which $k_B= c =\hbar= 1$. All quantities with dimension of energy are expressed in GeV when units are not specified.

\section{II. The model}

In~\cite{Bekenstein:1992pj} the conformal and disformal contributions to the physical metric were introduced and  generated by a scalar
field $\phi$. Defining a generic function $F(\phi, X, Y)$ by
\beq
\diff s^2 = \tilde g_{\mu \nu} \diff  x^\mu \diff x^\nu 
=g_{\mu \nu} \diff x^\mu \diff x^\nu F(\phi,X,Y)\,,
\label{Eq:ds2bis}
\eeq
with
\begin{equation}
    X = g^{\alpha \beta}\partial_\alpha \phi \partial_\beta \phi\,,
\quad \text{  } \quad
Y = \frac{\partial_\alpha \phi \diff x^\alpha \partial_\beta \phi \diff x^\beta}{g_{\alpha \beta} \diff x^\alpha \diff x^\beta},
\end{equation}
\noindent 
and
\beq
F=C(\phi, X) + D (\phi, X) Y\,.
\label{Eq:F}
\eeq
 The {\it physical} metric $\tilde g_{\mu \nu}$ then becomes
\beq
\tilde g_{\mu \nu} = 
C(\phi,X) g_{\mu \nu} 
+ D(\phi,X) \partial_\mu \phi \partial_\nu \phi\,.
\label{Eq:gtilde}
\eeq
The expression (\ref{Eq:gtilde}) contains a
{\it conformal} and a {\it disformal} transformation between the two metrics $g_{\mu \nu}$ and $\tilde g_{\mu \nu}$ induced 
by $C$ and $D$ respectively. The disformal coupling has been studied extensively in \cite{Dusoye:2020wom,Brax:2016kin} at the cosmological level and \cite{Trojanowski:2020xza,us} for WIMP and FIMP scenarii of dark matter respectively.
In both cases, a $Z_2$ symmetry was implicitly introduced to ensure the stability of the DM candidate. 
A common parametrization of the $C$ and $D$ functions is  given by \cite{Sakstein:2014aca}
\bea
&&
C(\phi,X) = e^{\alpha \frac{\phi}{M_P}}
= 1 + \frac{\alpha}{M_P} \phi
+{\cal O}\left(\frac{\phi^2}{M_P^2}\right)\,
\\
&&
D(\phi,X) = \frac{d}{M^4_P}e^{\beta \frac{\phi}{M_P}} 
= \frac{d}{M_P^4}+ \frac{\beta d}{M^5_P}\phi + {\cal O}\left(\frac{\phi^2}{M_P^6}\right)\,,
\nonumber
\eea
generating at the first order, the physical metric
\begin{equation}
\tilde g_{\mu \nu}\, = \, g_{\mu \nu} + \alpha\frac{\phi}{M_P} g_{\mu \nu}
+\frac{d}{M^4_P}\partial_\mu \phi \partial_\nu \phi \, = \, g_{\mu \nu} + \delta g_{\mu \nu}\,,
\end{equation}
with
\begin{equation}
\delta g_{\mu \nu}=\alpha \frac{\phi}{M_P} g_{\mu \nu}
+\frac{d}{M^4_P}\partial_\mu \phi \partial_\nu \phi\,.
\end{equation}
From now on, we will consider the phenomenology induced at the first order of perturbation theory in $\delta g_{\mu\nu}$. The highest temperature in the plasma $T_\text{max}$ 
being much lower than $M_P$ (even below the inflaton mass $m_\Phi \simeq 3 \times 10^{13}$ GeV \cite{Garcia:2017tuj,Barman:2021tgt,GKMO1}), 
the disformal part of the metric generates terms 
 $\lesssim \frac{T_\text{max}^2}{M_P^2}\frac{\phi^2}{M_P^2}$, which are expected to have little influence on the dark matter phenomenology\footnote{For a specific analysis of the disformal term in the dark matter production in the earliest stage of the Universe, see \cite{us}.} 
for reasonable values of $\alpha$. The perturbative part of the metric $\delta g_{\mu \nu}$ induces couplings to the Standard Model of the form
\bea
\delta {\cal S}_{\rm SM}\, =&&\, \frac{1}{2}\int \diff^4x \sqrt{-g} T^{{\rm SM}}_{\mu \nu} \delta g^{\mu \nu}\,,
\nonumber
\\ \, =&&\,-\frac{\alpha}{2}\frac{\phi}{M_P}\int \diff ^4 x  \sqrt{-g}T_{\mu \nu}^{{\rm SM}}g^{\mu \nu}\,,
\label{eq:deltaS_SM}
\eea
where $T_{\mu \nu}^{{\rm SM}}$ is the total SM energy-momentum tensor \footnote{Details can be found in the Appendix.} that can be expressed as
\begin{equation}
    T^{{\rm SM}}_{\mu \nu} \, = \,  \sum_{i=0,1/2,1} T^{i}_{\mu \nu} - g_{\mu \nu} {\cal L}_{\rm int}\,,
    \label{eq:TSM}
\end{equation}
where $T^{i}_{\mu \nu}$ represents individual contributions from SM particles of spin ($i=0,1/2,1$) fields to the total energy momentum tensor, as given by
\bea
&&
T_{\mu \nu}^{0} =
2(D_\mu H^\dagger)(D_\nu H) - g_{\mu \nu}
\left[D^\alpha H^\dagger D_\alpha H  \right] \,,
\nonumber
\\
&&
T_{\mu \nu}^{1/2} = \sum_{\psi}
\frac{i}{4}
\Big[
\bar \psi \gamma_\mu \lrD_\nu \psi
+\bar \psi \gamma_\nu \lrD_\mu \psi \Big] -g_{\mu \nu}\Big[\frac{i}{2}
\bar \psi \gamma^\alpha \lrD_\alpha \psi\Big]\,,
\nonumber
\\
&&
T_{\mu \nu}^1 = \sum_{A_\mu}   \frac{1}{4}  g_{\mu \nu} F^{\alpha \beta} F_{\alpha \beta} - F_\mu^{\;\alpha} F_{\nu \alpha} \,,
\eea
where $H$ is the SM Higgs doublet, $\psi$ represents   SM fermion  and $A_\mu$ a SM gauge field $A_\mu$ with corresponding field strength tensor $F_{\mu \nu}$. The sums are performed over all SM fields. $D_\mu\equiv \del_\mu -i q_a g_a A_\mu$ is the covariant derivative with respect to an appropriate $A_\mu$ with a gauge coupling $g_a$ and charge $q_a$. $\lrD_\mu\equiv\rD_\mu-\lD_\mu$ with $\rD_\mu \psi=\del_\mu \psi-ig_a q_a A_\mu \psi$ and $\bar \psi\lD_\mu = \del_\mu\bar \psi+i g_a q_a \bar \psi A_\mu$.  Non-abelian representation indices are omitted for clarity but the generalization is straightforward. ${\cal L}_{\rm int}$ is the contribution to the Lagrangian defined as
\beq
{\cal L}_{\rm int} =
-V(H)+{\cal L}_{\text{Y}}\,,
\label{eq:Lint}
\eeq
with the Yukawa Lagrangian being
\begin{equation}
{\cal L}_{\text{Y}}=-\left[y_t \bar Q_L \tilde H t_R+y_b\bar Q_L H b_R+y_\ell \bar L_L H \ell_R + \text{h.c.}\right]\,,
\label{eq:LYukawa}
\end{equation}
where $\ell=e,\mu,\tau$ denotes SM leptons with corresponding $SU(2)_L$ doublet $L_L=(\nu_L \; \ell_L)^T$, $\nu_L$ is the SM left-handed neutrino state of flavour $\ell$. Only the third generation of SM quarks is represented, i.e. top ($t$) and bottom ($b$) quarks, with corresponding $SU(2)_L$ doublet $Q_L=(t_L \; b_L)^T$ and flavour indices are omitted for clarity. $V(H)$ is the usual SM Higgs scalar potential parametrized as
\beq
V(H)= - \mu^2 |H|^2 + \lambda |H|^4\,.
\label{eq:V(H)}
\eeq
The total Lagrangian can be expressed as
\begin{multline}
    {\cal L}= {\cal L}_{\rm SM} -\frac{\alpha}{2}\frac{\phi}{M_P} \Big[   4 V(H)-2 D_\mu H^\dagger D^\mu H- 4 \mathcal{L}_\text{Y} \\ -\sum_\psi \frac{3i}{2} \big(\bar \psi  \gamma_\mu D^\mu \psi - D^\mu \bar \psi \gamma_\mu \psi \big) \Big]
       \label{eq:Lag_tot}
\end{multline}
with ${\cal L}_{\rm SM}$ being the total SM Lagrangian.
Notice that on mass shell, only the spin 0 fields
should be taken into account at temperatures above the electroweak breaking phases, 
where all the standard model particles are massless\footnote{Thermal masses are generated at higher order, but will be subdominant to the scattering involving Higgs fields.}. 
Noticing that the fermions and gauge bosons of the Standard Model are massless
at the scales of interest, nullifying the trace of their stress-energy tensor, only the coupling of $\phi$ to the Higgs field will survive. Moreover, at temperatures much above the electroweak scale,
$\frac{\mu^2}{T^2},~\frac{|H|^2}{T^2}\ll 1$, meaning that we can also neglect the $V(H)$ term in Eq.~(\ref{eq:Lag_tot}). Finally, even if the field $\phi$ is clearly unstable, we will see that it can 
still be a viable dark matter candidate, with a lifetime much larger than the age of the Universe if the Beyond Standard Model (BSM) scale $\Lambda\equiv \frac{M_P}{\alpha} \gtrsim 10^{14}$ GeV. But what is even more remarkable is that its coupling to the Higgs field through the Higgs-kinetic term
ensures a sufficient amount of dark matter to fulfill the cosmological constraints
thanks to processes involving the top quark whose large Yukawa coupling to the Higgs
field compensates for the Planck scale suppression.

\section{III. Dark matter phenomenology}

\subsection{Relic abundance constraint}
\begin{figure*}[ht]
\centering
\includegraphics[width=3.in]{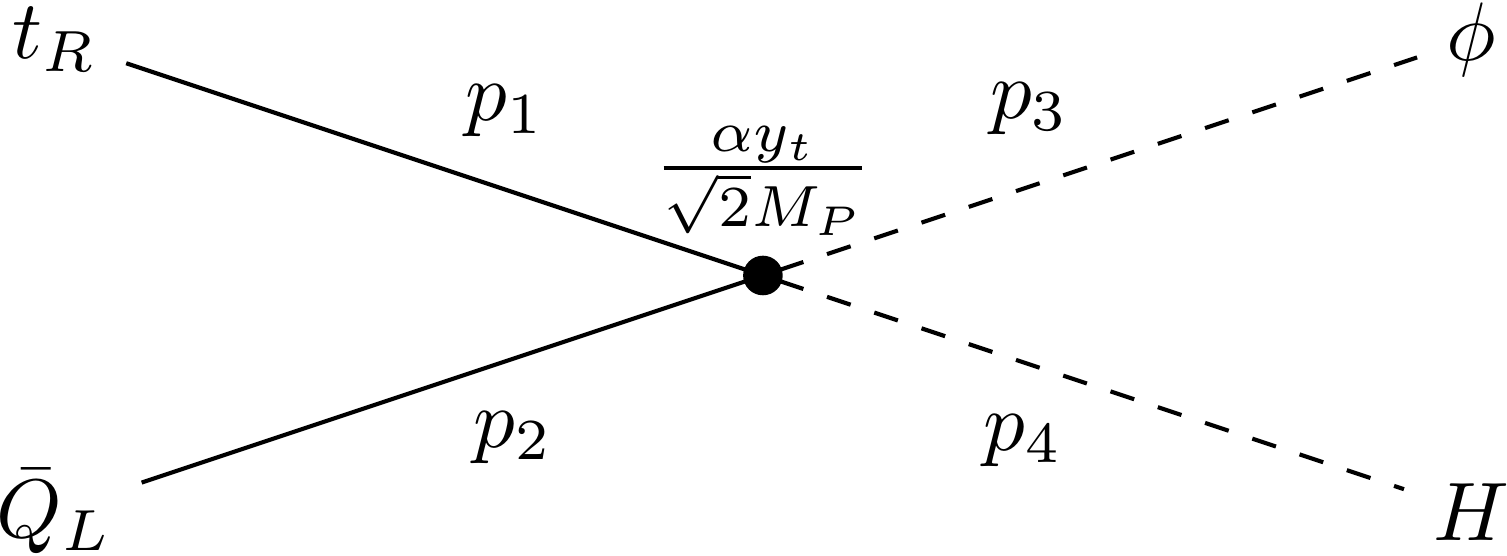}
~~~
\includegraphics[width=3.in]{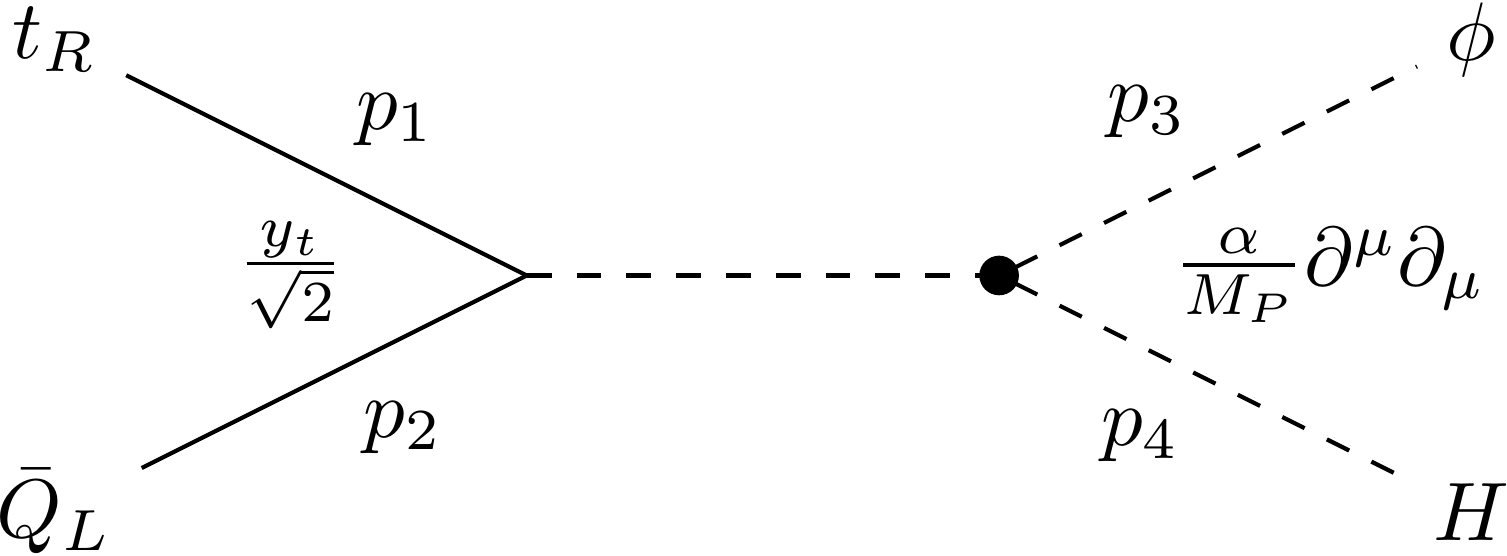}
\caption{\em \small Dominant scattering processes contributing to the population of the dark matter candidate $\phi$. In addition to $t_R\bar Q_L\to \phi H$ shown here, $t_R H (\bar Q_L H)\to Q_L\phi (\bar t_R \phi)$ also exist, which corresponds to taking the time direction from bottom to top in these diagrams.}

\label{Fig:feynman}
\end{figure*}

In our setup the dark matter is produced at high temperatures, before the Electro-Weak Symmetry Breaking (EWSB). Based on the Lagrangian of Eq.~(\ref{eq:Lag_tot}), many production modes contribute to this process, however one can understand that the dominant processes will be the ones involving the top and bottom quarks, whose couplings to the Higgs field are the largest ones. Such processes are represented in Fig.~\ref{Fig:feynman}.
It is therefore clear that the equivalent processes with gauge bosons and a fortiori other types of quarks or leptons in the initial state 
will be suppressed  compared to the top-quark Yukawa coupling $y_t$ by a factor
$\propto c_i^2/y_t^2$ where $c_i$ represents  any dimensionless coupling (i.e. Yukawa, gauge or scalar-potential couplings) involving the scattering species $i$.
\footnote{There also exist anomaly induced couplings of the form $\phi F^{\mu\nu}F_{\mu\nu}$ leading to, for instance, gluon-gluon to gluon-$\phi$ channels. Such diagrams diverge in the infrared region, such as collinear regions in phase space. To regulate them, we may utilize the thermal masses of the involved gauge bosons, which is however largely beyond the scope of our paper. Nevertheless, the reaction rate should be proportional to $1/M_P^2$ as in the case of the Yukawa interaction contributions which we compute.}
At high temperatures, where electroweak symmetry is restored, the Higgs doublet can be parametrized as
\begin{equation}
    H=\begin{pmatrix}H^+\\H^0\end{pmatrix}=\frac{1}{\sqrt{2}}\begin{pmatrix}h_1+ih_2\\h_3+ih_4\end{pmatrix}\,,
\end{equation}
in terms of complex scalars $H^+$, $H^0$ or in terms of real scalars $h_i$ with $i=1,2,3,4$.
The most relevant terms of the Yukawa Lagrangian above EWSB are
\begin{equation}
{\cal L}_{\rm Y}\supset-  y_t(     \bar t _L t_R H^{0\,\dagger}-\bar b_L t_R H^- +\text{h.c.})\,,
\label{eq:Lag_Yuk_above_EWSB}
\end{equation}
where $H^-=(H^+)^*$. 
The dark matter abundance produced in processes such as the one depicted in 
Fig.~\ref{Fig:feynman} can be estimated by solving the Boltzmann equation 
\begin{equation}
\frac{\diff n_\phi}{\diff t} + 3 H n_\phi(t) = R(t)\,,
\end{equation}
with $R(t)$ being the time-dependent DM production rate per unit of volume and time. The Boltzmann equation can be expressed in term of the SM plasma temperature $T$ as
\begin{equation}
    \frac{\diff Y_\phi}{\diff T} =-\frac{R(T)}{H(T)T^4}\,,
\label{Eq:boltzmann}
\end{equation}
where the quantity $Y_\phi(T)\equiv  n_\phi(T)/T^3$ is proportional to the DM number per comoving volume. $H\equiv \dot a/a=(g_T \pi^2/90)^{1/2}T^2/M_P$ 
is the Hubble expansion parameter.\footnote{We will consider the reduced Planck mass $M_P=2.4\times 10^{18}$ GeV throughout our study and $g_T$ the 
effective relativistic degrees of freedom at a temperature $T$.} The quantity $Y_\phi(T)$ becomes constant once the dark matter production 
is frozen by the expansion
($R(T)\ll T^3 H $)\footnote{By neglecting the temperature evolution of the effective relativistic degrees of freedom.}.
The Boltzmann equation, in the earliest phase of the Universe, taking into account non-instantaneous reheating and/or 
non-instantaneous thermalization has been studied extensively in the literature
recently (see  \cite{Garcia:2017tuj,Barman:2021tgt,GKMO1} 
and \cite{Reheating,Elahi:2014fsa} for instance).
In order to solve Eq.~(\ref{Eq:boltzmann}), one needs to compute the average production rate 
$R(T)$ which can be expressed for processes labeled by $1 + 2 \rightarrow 3 + 4$, where $1,2$ and $3,4$ denote respectively the initial and final states, by 
\beq
R(T) = \frac{1}{1024 \pi^6}\int f_1 f_2 E_1 \diff E_1 E_2 \diff E_2 \diff \cos \theta_{12}\int |{\cal M}|^2 \diff \Omega_{13}.
\nonumber
\eeq

\noindent
where $E_i$ denote the energy of particle $i=1,2,3,4$ and 
\beq
f_{i}= \frac{1}{e^{E_i/T}\pm 1}\,.
\eeq
represent the (thermal) distributions of the incoming particles. Using the Lagrangian of Eq.~(\ref{eq:Lag_Yuk_above_EWSB}), we obtain
\beq
R^{\rm scat}(T) = \frac{567 \alpha^2 y_t^2 \zeta(3)^2}{256 \pi^5}
\frac{T^6}{M_P^2}\equiv \beta \frac{T^6}{M_P^2}\,,
\eeq
\noindent
with $\beta\simeq 0.01\,\alpha^2$, which gives, after integrating   Eq.~(\ref{Eq:boltzmann}) and at low temperature
\beq
Y_\phi^{\rm scat} = \beta\sqrt{\frac{90}{g_T\pi^2}}\frac{T_{\rm RH}}{M_P}\,,
\eeq
\noindent
with $T_{\rm RH}$ being the reheating temperature. The corresponding relic abundance at the present time
\begin{equation}
\Omega_\phi^{\rm scat} h^2 =\frac{n^{\rm scat}_\phi(T_0) m_\phi}{\rho_c^0/h^2}
\simeq 1.6 \times 10^{8} Y^{\rm scat }_\phi
\left(\frac{g_0}{g_\text{RH}} \right)
\left(\frac{m_\phi}{1~\mrm{GeV}} \right)\,,
\label{Eq:omega}
\end{equation}
\noindent
where $\rho_c^0/h^2 = 1.05 \times 10^{-5} ~\mrm{GeV \,cm^{-3}}$ is the present critical density and $g_i$ denotes the effective number of degrees of freedom at temperature $T_i$.\footnote{With $g_0=3.91$, $g_\text{RH}=106.75$ for reheating temperatures larger than the top-quark mass $T_\text{RH}>m_t$ in the Standard Model.} The present relic abundance can be expressed as
\bea
&&
\frac{\Omega_\phi^{\rm scat} h^2}{0.1}\simeq  
\left( \frac{\alpha}{37} \right)^2
\left( \frac{T_{\rm RH}}{10^{10}~\mrm{GeV}} \right)
\left( \frac{m_\phi}{1~\mrm{GeV}}\right)\,.
\label{Eq:omegabis}
\eea
Note that the DM mass $m_\phi$ is a free parameter and defined in the canonical form in the Einstein frame. More details are provided in the Appendices.

\noindent

\subsection{Production from inflaton decay}

In general, interactions between the inflaton and dark matter may also arise from the trace of the inflaton stress-energy tensor.
Suppose that the inflaton $\Phi$ has a coupling to the SM Higgs field to achieve reheating, given by ${\cal L}\supset -\mu_\Phi \Phi |H|^2$. Including such a term in Eq.~(\ref{eq:Lint}), we obtain
\bea
{\cal L}_\phi \supset 2\alpha\frac{\mu_\Phi}{M_P}\phi\Phi|H|^2
\label{eq:Lphi}
\eea
through the trace of the corresponding energy--momentum tensor,\footnote{ The form of the inflaton couplings to $\phi$ depends on in which frame, i.e., Jordan or Einstein, the inflaton sector is introduced. Here, we assume that Eq.~(\ref{eq:Lphi}) is the only coupling between $\phi$ and $\Phi$ in the Einstein frame. The rest of the inflaton potential in the same frame is assumed to be independent on $\phi$. Additional couplings may be present and may be required to avoid the generation of isocurvature perturbations during inflation. Such a  possible coupling is also considered in Appendix B to avoid creating large isocurvature perturbations.} generating a possible inflaton decay into $\phi H H$  given by
\bea
\Gamma^\Phi_{\phi HH} = \frac{\alpha^2\mu_\Phi^2 m_\Phi}{128\pi^3M_P^2}.
\eea
The dominant decay channel of the inflaton is into a pair of  SM Higgs bosons, whose decay width is given by
\bea
\Gamma^\Phi_{HH} = \frac{\mu_\Phi^2}{8\pi m_\Phi}.
\eea
The branching fraction of the single dark matter production is obtained by
\bea
B_R = \frac{\Gamma^\Phi_{\phi HH}}{\Gamma^\Phi_{HH}} = \left(\frac{\alpha}{4\pi}\right)^2\left(\frac{m_\Phi}{M_P}\right)^2.
\eea
The dark matter number density from the inflaton decay is then estimated as\footnote{ Here, we have taken into account the non-instantaneous reheating effect \cite{Kaneta:2019zgw}.}
\bea
\frac{n_\phi^{\rm dec}(T_{\rm RH})}{T_{\rm RH}^3} = B_R\times\frac{g_{\rm RH}\pi^2}{18}\frac{T_{\rm RH}}{m_\Phi}\,,
\eea
from which we obtain 
\begin{align}
\frac{\Omega_\phi^{\rm dec}h^2}{0.1} \simeq \left(\frac{\alpha}{37}\right)^2\left(\frac{m_\Phi}{3\times10^{13}~{\rm GeV}}\right)\left(\frac{T_{\rm RH}}{10^{10}~{\rm GeV}}\right)\frac{m_\phi}{660~{\rm GeV}}\,.
\label{Eq:omegadecay}
\end{align}
Comparing Eqs.~(\ref{Eq:omegabis}) and (\ref{Eq:omegadecay}), we clearly see that the contribution from the inflaton decay is always subdominant with respect to to the scattering processes. In fact, this is not an usual feature of FIMP produced after the inflationary stage. In the case of conformal dark matter, the coupling of $\phi$ to the inflaton $\Phi$ is {\it extremely} reduced by a scale $B_R\propto (\frac{m_\Phi}{M_P})^2\simeq 10^{-10}$ whereas it was shown in \cite{Kaneta:2019zgw} that a branching ratio of $\sim 10^{-6}$ is necessary to produce a cosmologically viable 1 GeV dark matter candidate at $\trh=10^{10}$ GeV.  

\subsection{Lifetime constraint}
The first condition for $\phi$ to be a good dark matter candidate is that it should have  a sufficiently
long lifetime, at least of the order of the age of the Universe. There are also constraints coming from neutrino or gamma-ray observations \cite{Mambrini:2015sia}.  Assuming that $m_\phi$ lies below the electroweak symmetry breaking scale, the Higgs doublet can be parametrized in unitary gauge by
\begin{equation}
    H=\dfrac{1}{\sqrt{2}}\left( \begin{array}{c}
        0 \\
        v_h+ h
    \end{array} \right)\,,
\end{equation}
where $h$ denotes the real SM physical scalar degree of freedom below the EWSB scale and $v_h\simeq 246$ GeV being the Higgs vev\footnote{This parametrization corresponds to $H^+=0$ and $H^0=(v_h+h)/\sqrt{2}$ or $h_{1,2,4}=0$ from the parametrization used to described physics above the EWSB scale.}. Below the EWSB scale, the DM candidate $\phi$ still couples to the entire SM spectrum via the Lagrangian already given in Eq.~(\ref{eq:Lag_tot}). However for on-shell fermionic SM states, by applying the equations-of-motion to Eq.~(\ref{eq:Lag_tot}), interactions between the DM and a pair of SM particles can be described by the following terms
\begin{multline}
 \mathcal{L}\supset   \alpha  \frac{\phi}{M_P}  \Big(\dfrac{m_Z^2}{2} Z^\mu Z_\mu+m_W^2 W^{- \mu} W^+_\mu \\- \dfrac{m_\psi}{2} \bar \psi \psi  +\dfrac{1}{2}\partial_\mu h \partial^\mu h  - m_h^2 h^2 \Big)\,,
     \label{eq:LagbelowEWSB}
\end{multline}
where $Z,W$ are the massive weak gauge bosons.
Since the coupling of the DM candidate to on-shell SM fermions occurs via the Yukawa couplings, at first sight we could expect the neutrinos to decouple from the DM candidate as no such Yukawa terms exist for neutrinos in the SM.\footnote{We remind the reader that to this day, neutrinos possess non-vanishing masses however their generation mechanism have not been identified yet.} However, notice that interaction terms with fermions originate from the DM coupling to the total stress-energy tensor via Eq.~(\ref{eq:deltaS_SM}), which itself contains kinetic terms, therefore regardless of the neutrino-mass generation mechanism, such couplings with neutrinos should always be present. \par \medskip

{ There are also coupling terms coming from the  trace anomaly and originating from triangle diagrams where states of masses less than $m_\phi$ run in the loop.}
The appearance of such terms can be understood by recalling that in Eq.~(\ref{eq:deltaS_SM}) the terms such as $F^a_{\mu\nu}F^{a\mu\nu}$ is proportional to $d-4$, and thus they vanish when $d=4$ at tree level, whereas at loop level, wave function renormalization factor contains the terms proportional to $1/(d-4)$, and thus finite terms remain.
There also exist the contributions from the Yukawa and Higgs quartic coupling beta functions, which are however irrelevant for our discussion.
One can calculate the dark matter coupling to the gauge bosons $V$:
\bea
{\cal L}_{\phi VV} = 
\frac{\alpha\phi}{16\pi M_P}b_a \alpha_a F^a_{\mu\nu}F^{a\mu\nu}
\label{eq:phi-V-V}
,
\eea
where $b_a=b_Y, b_2, b_3$ and $\alpha_a\equiv g_a^2/4\pi=\alpha_Y, \alpha_2, \alpha_s$ are the beta-function coefficients and gauge couplings of U(1)$_Y$, SU(2)$_L$, and SU(3)$_C$, respectively. 
The gauge field strength is respectively represented by $F^a_{\mu\nu}=G^a_{\mu\nu}, W^a_{\mu\nu}, B_{\mu\nu}$ for SU(3)$_C$ SU(2)$_L$, and U(1)$_Y$ gauge fields.
Notice that for the SU(2)$_L\times$U(1)$_Y$ piece, once the electroweak symmetry is broken, the gauge fields are transformed into  mass eigenstates which is however not necessarily the basis where the photon ($A_\mu$) and the weak gauge bosons ($W^\pm_\mu$ and $Z_\mu$) are orthogonal in Eq.~(\ref{eq:phi-V-V}).
Indeed, we obtain
\begin{align}
{\cal L}_{\phi VV} \supset
&&
\frac{\alpha\phi}{M_P}\left(c_{WW} W^+_{\mu\nu}W^{-\mu\nu}
+c_{ZZ} Z^{\mu\nu}Z_{\mu\nu}\right.\nonumber\\ 
&&
\left.
+c_{Z\gamma}Z^{\mu\nu}A_{\mu\nu}
+c_{\gamma\gamma}A^{\mu\nu}A_{\mu\nu}
\right)\,,
\end{align}
with
\begin{align}
c_{ZZ}
\,
& =\,\frac{(b_2/t_W^2+b_Y t_W^2)\alpha_{\rm em}}{16\pi}\,, \; && c_{WW}  \,
=\,\frac{b_2\alpha_{\rm em}}{8\pi s_W^2}\,, \nonumber\\
c_{Z\gamma}
\,
& =\,\frac{(b_2/t_W-b_Yt_W)\alpha_{\rm em}}{8\pi}\,,  \; &&
c_{\gamma\gamma}\,
=\,\frac{(b_2+b_Y)\alpha_{\rm em}}{16\pi}\,,\nonumber
\end{align}
where the field strengths are assumed not to include self-interacting terms, and $\alpha_{\rm em}\equiv\alpha_2/s_W^2=\frac{e^2}{4 \pi}=\frac{1}{137} $  is the fine structure constant, $s_W\equiv\sin\theta_W, t_W\equiv \tan\theta_W$ with $\theta_W$ being the weak mixing angle. The corresponding decay rate are given by 
\begin{equation}
\Gamma^\phi_{hh}= 
\frac{\alpha^2 m_\phi^3}{128 \pi M_P^2}
\simeq 6\times 10^{-28} 
\left(\frac{\alpha}{37} \right)^2
\left( \frac{m
_\phi}{1~\mrm{TeV}}\right)^3\mrm{GeV}\,,
\end{equation}
and
\begin{align}
\Gamma^\phi_{\bar \psi \psi}\,&=\, c_\psi
\frac{\alpha^2 m_\phi m_\psi^2}{32 \pi M_P^2}
\nonumber\\
\,&=\, 2  \times 10^{-36}\, \mrm{GeV}\,
c_\psi~
\left( \frac{\alpha}{37}\right)^2
\left( \frac{m_\psi}{1~\rm{GeV}}\right)^2
\left(\frac{m_\phi}{1~\rm{GeV}} \right)\,,
\end{align}
which are effective only when $m_\phi$ is large enough to allow for the Higgs and fermion productions. The coupling to photons implies the following decay rate
\begin{align}
\Gamma^\phi_{\gamma\gamma}\,= &\,\frac{c_{\gamma\gamma}^2\alpha^2}{4\pi}\frac{m_\phi^3}{M_P^2}\nonumber\,,
\\
\,\simeq &\,
\frac{c_{\gamma\gamma}^2}{c_{ZZ}^2}\Gamma^\phi_{ZZ}
\simeq
\frac{2c_{\gamma\gamma}^2}{c_{Z\gamma}^2}\Gamma^\phi_{Z\gamma}
\simeq
\frac{c_{\gamma\gamma}^2}{2c_{WW}^2}\Gamma^\phi_{WW}
\simeq
\frac{c_{\gamma\gamma}^2}{c_gc_{gg}^2}\Gamma^\phi_{gg}
\nonumber\,,
\\
=&\,
2.3\times10^{-31}{\rm GeV}
\left( \frac{\alpha}{130}\right)^2
\left( \frac{c_{\gamma\gamma}}{10^{-3}}\right)^2
\left( \frac{m_\phi}{1~{\rm TeV}}\right)^{ 3},
\end{align}
where $c_{gg}\equiv b_3\alpha_s/16\pi$. Similarly $c_\psi$ and $c_g=8$ are the  dimensions of the $SU(3)$ representations for the fermion $\psi$ and  the gluons. Notice that when $m_\phi$ is smaller than the electron mass, dark matter does not decay through the anomaly since $b_a=0$ and other processes contribute to the decay into photons.  We then obtain numerically
\begin{align}
\Gamma^\phi_{hh}\,=&\, 6 \times 10^{-31}
\left(\frac{\Omega_\phi h^2}{0.1} \right)
\left(\frac{10^{10}~\rm{GeV}}{T_{\rm RH}} \right)
\left(\frac{m_\phi}{1~\rm{TeV}} \right)^2
\rm{GeV}
\nonumber
\\
\Gamma^\phi_{\bar \psi \psi}\,=&\,2 \times 10^{-36}
c_\psi
\left(\frac{\Omega_\phi h^2}{0.1} \right)
\left( \frac{m_\psi}{1~\rm{GeV}}\right)^2
\left( \frac{10^{10}~\rm{GeV}}{T_{\rm RH}}\right)
\rm{GeV}
\nonumber
\\
&&
\label{Eq:gammaf}
\end{align}
At low energy, in the unitary gauge the Higgs degree of freedom is a single real scalar 
and not a complex doublet. Remembering that the age of the Universe corresponds to a rate such that $\Gamma_{\rm Universe}^{-1} \simeq 10^{42} ~ \mrm{GeV^{-1}}$,  one concludes that the only tree-level decay
that may break the stability of DM over the age of the Universe is the neutrino channel\footnote{In fact, even if decays into electrons generate a lifetime larger than the age of the Universe, the electromagnetic nature of the final state imposes constraints of the order
$\Gamma^\phi \lesssim 10^{-51}$ GeV, due to CMB
constraints \cite{Mambrini:2015sia}. We see from Eq.(\ref{Eq:gammaf}) that this is far from being respected by $\Gamma^\phi_{ee}$ if we want $\Omega h^2 \simeq 0.1$.}  $\phi \rightarrow \nu \nu$, with $m_\nu \lesssim 0.05$ eV. We then obtain the relation
\begin{equation}
\Gamma^\phi_{\nu \nu} = 5 \times 10^{-57}
\left(\frac{\Omega_\phi h^2}{0.1} \right)
\left( \frac{m_\nu}{0.05~\rm{eV}}\right)^2
\left( \frac{10^{10}}{T_{\rm RH}}\right)
\rm{GeV}.
\label{Eq:widthneutrino}
\end{equation}
\noindent
A quick look at the expression above is sufficient to understand that the cosmological constraints $\Omega_\phi h^2\simeq0.1$ can be satisfied while still preserving a sufficiently long lifetime for $\phi$ as long as $m_\phi \lesssim 2 ~m_{e}$.

\begin{figure*}[ht]
\centering
\includegraphics[width=2.in]{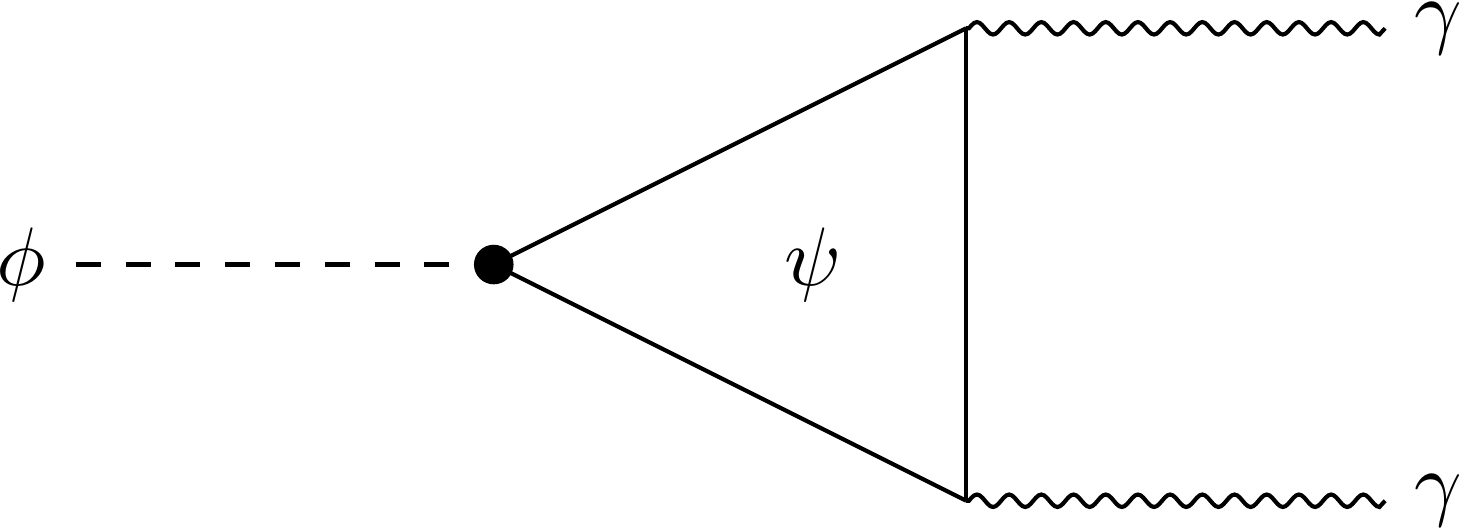}
\includegraphics[width=2.in]{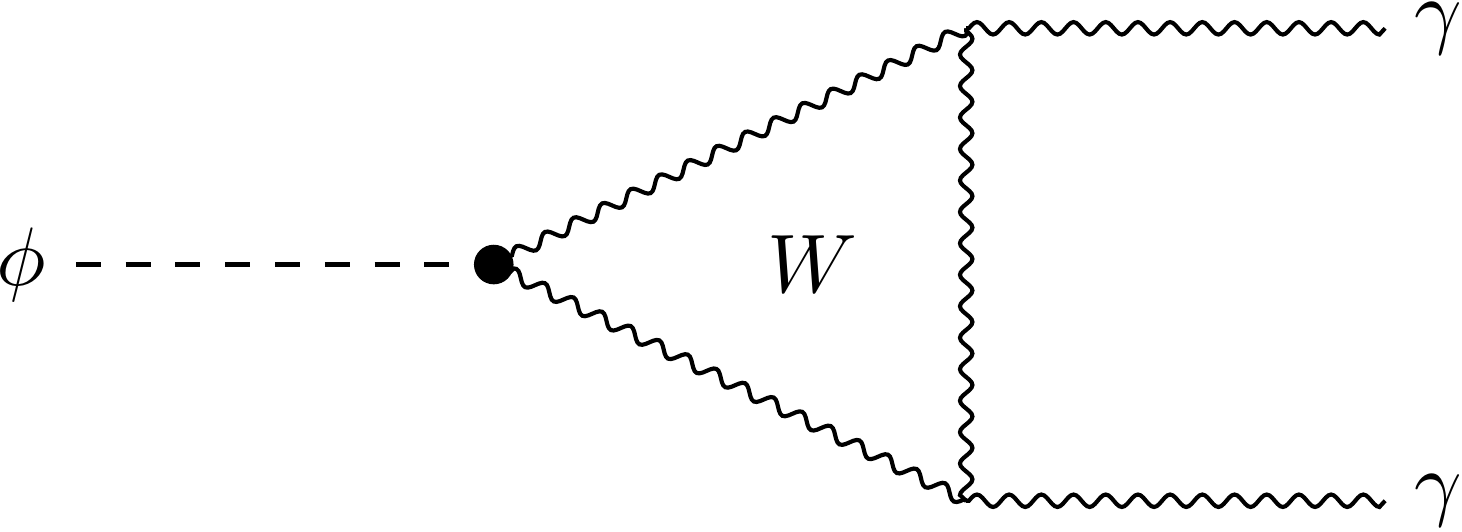}
\includegraphics[width=2.in]{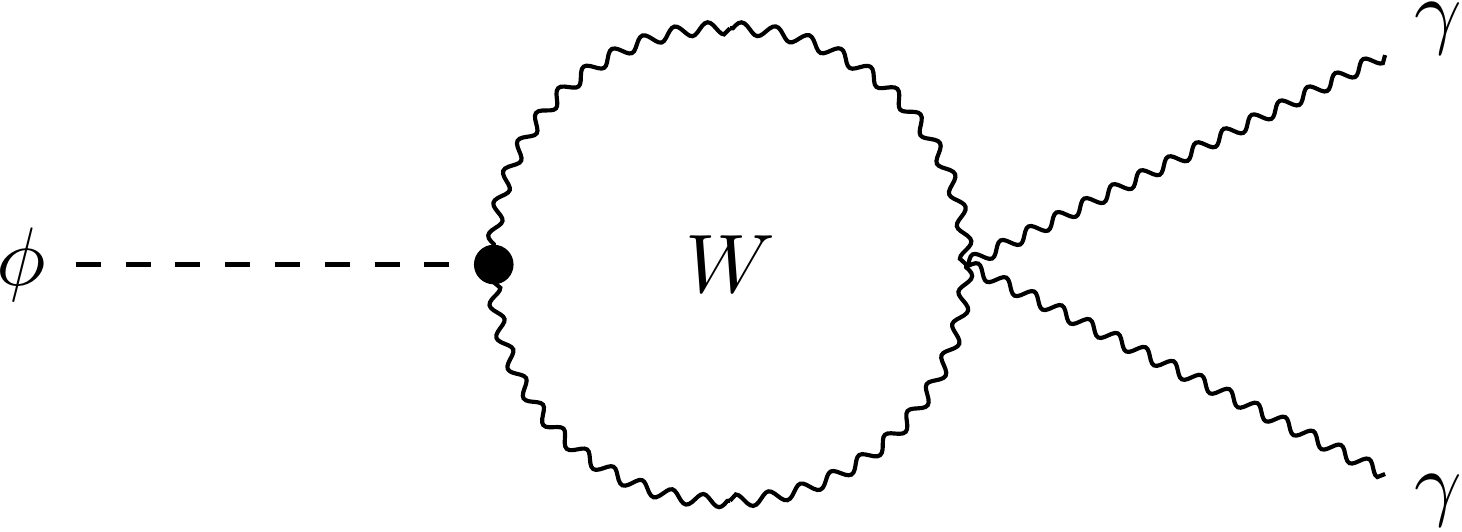}
\caption{\em \small Loop diagrams that contributes $\phi\to\gamma\gamma$, where we assume the unitary gauge for the W boson loops.}
\label{Fig:loop}
\end{figure*}

\begin{figure}[ht]
\centering
\includegraphics[width=3.5in]{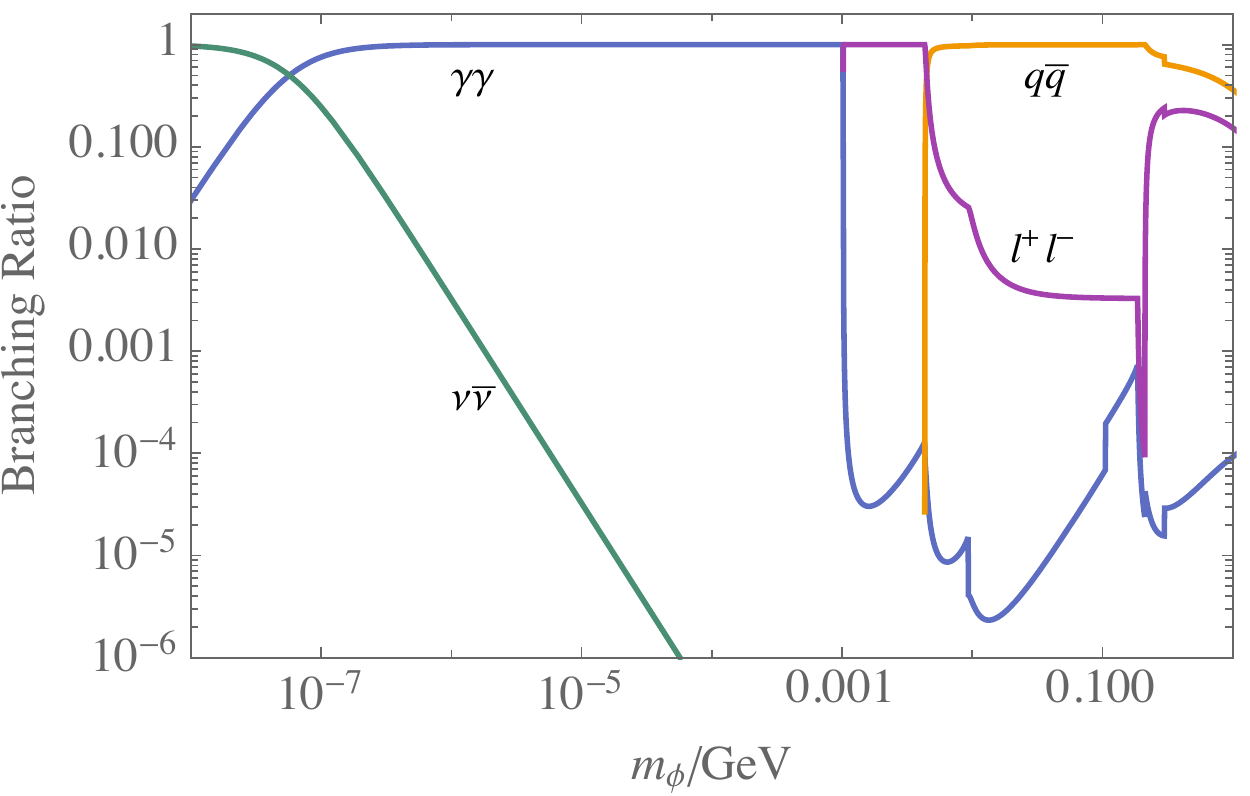}
\caption{\em \small Branching ratios of the decay of the conformal dark matter.}
\label{Fig:BR}
\end{figure}

Loop-induced processes leading $\phi \rightarrow \gamma \gamma$ are also relevant,  whose diagrams are shown in Fig. \ref{Fig:loop}.
Indeed, 
when $m_\phi< m_e$,  the decay through the trace anomaly does not happen, as the beta functions cancel as we have already seen.
Instead, $\phi\to\gamma\gamma$ takes place through charged fermion and $W$ boson loops, whose decay width is given by
\beq
\Gamma^\phi_{\gamma\gamma} =\frac{\alpha^2\alpha_{\rm em}^2}{1024\pi^3}
\left|\sum_\psi N_c^\psi Q_\psi^2 A_{1/2}(\tau_\psi)+ A_1(\tau_W) \right|^2
\frac{m_\phi^3}{M_P^2}
\eeq
with $\tau_i \equiv m_\phi^2/4m_i^2$. $Q_\psi$ is the electric charge of a fermion $\psi$ with corresponding color factor $N_c^\psi$. We defined
\bea
&&
A_{1/2}(\tau)=\frac{2}{\tau^2}[\tau + (\tau-1)\arcsin^2(\sqrt{\tau})]\,,
\nonumber
\\
&&
A_{1}(\tau)=-\frac{1}{\tau^2}[2 \tau^2+3\tau+3(2 \tau-1)\arcsin^2(\sqrt{\tau})]\,.
\nonumber
\eea
This expression has been derived using the full Lagrangian of Eq.~(\ref{eq:Lag_tot}) which includes notably fermionic derivative terms, not appearing in Eq.~(\ref{eq:LagbelowEWSB}). With the hypothesis $m_\nu \ll m_\phi \ll m_e$ we obtain
\beq
\Gamma_{\gamma \gamma}^\phi = \frac{121~ \alpha^2 \alpha_{\rm em}^2}{9216 \pi^3}
 \frac{m_\phi^3}{M_P^2}\simeq 
 3.9\times10^{-45}\alpha^2\left(\frac{m_\phi}{1~\rm{GeV}}\right)^3\rm{GeV}
 \,.
 \label{Eq:gammagamma}
\eeq
\noindent
which gives, combining with Eq.~(\ref{Eq:omegabis})
\bea
\Gamma^\phi_{\gamma\gamma}=5.3 \times10^{-52}~
\left(\frac{\Omega_\phi h^2}{0.1} \right)
\left( \frac{10^{10}}{\trh}   \right)
\left(\frac{m_\phi}{10~{\rm keV}}\right)^2\,.
\label{Eq:widthgamma}
\eea
\noindent
and finally putting together  Eqs.~(\ref{Eq:widthneutrino}) and (\ref{Eq:widthgamma}), we obtain

\bea
&&
\tau_\phi\simeq \left(\frac{0.1}{\Omega_\phi h^2}\right)\frac{\trh}{10^{10}}
\left(\frac{10~\mrm{keV}}{m_\phi}\right)^2
\frac{1.3\times 10^{27} ~\rm{s}}{1+10^{-5}
\left(\frac{m_\nu}{0.05~\rm{eV}}
\frac{10~\rm{keV}}{m_\phi}\right)^2}
\nonumber
\\
&&
\simeq \left(\frac{0.1}{\Omega_\phi h^2}\right)
\frac{\trh}{10^{10}}
\left(\frac{10~\mrm{keV}}{m_\phi}\right)^2
1.3\times 10^{27} ~\rm{s.}
\label{Eq:lifetime}
\eea
\noindent
The last equality shows us that,
taking into account the cosmological limit $m_\nu \lesssim 0.05$ eV, the $\gamma \gamma$ final 
state is always the dominant channel process for $m_\phi\gtrsim 10$ eV. This lower bound on $m_\phi$ is already excluded by the Lyman-$\alpha$ constraint as we will see in the next section where in fact we will require that $m_\phi\gtrsim 3.9 $ keV. We illustrate this feature in Fig.~\ref{Fig:BR}, where we can clearly see that the $\gamma\gamma$ channel dominates when $m_\phi<2m_e$,
while the $e^+e^-$ channel dominates until the $q\bar q$ channel opens for $m_\phi>2m_u$.

\subsection{Lyman-$\alpha$ constraints}

Dark matter candidates with a non-negligible contribution to the cosmological background pressure can alter the matter power spectrum of density fluctuations by erasing overdensities on small physical scales. As a  result this introduces   a cutoff at large Fourier wavenumbers compared to the power spectrum expected within the $\Lambda$CDM cosmology. Absorption lines around $\sim 100\, $nm of light emitted by distant quasars at redshifts $z\sim2-6$ by the neutral Hydrogen of the intergalactic medium, known as  the Lyman-$\alpha$ forest, allow to probe the matter power spectrum on scales $k\sim (0.1-10)\, h\,\text{Mpc}^{-1}$. For a given dark matter phase space distribution, the Lyman-$\alpha$ forest can be used to set a bound on the DM mass or alternatively the equation-of-state parameter of such species. The Lyman-$\alpha$ bound is typically given in terms of a mass for Warm Dark Matter (WDM)~\cite{Narayanan:2000tp,Viel:2005qj,Viel:2013fqw,Baur:2015jsy,Irsic:2017ixq,Palanque-Delabrouille:2019iyz,Garzilli:2019qki},\footnote{Here, WDM is defined as DM species with a thermal-like distribution such as Fermi-Dirac distribution characterized by a parameter $T_{\rm WDM}$ playing the role of temperature, which is fixed by the requirement of reproducing the total observed dark matter density, for a given mass.}
\beq \label{bound1}
m_{\rm WDM}\;\gtrsim\; m_{\rm WDM}^{\text{Ly}-\alpha}\; =\; (1.9-5.3)~\text{keV at 95\% C.L.} \,,
\eeq
whose precise value depends on the specific analysis. 
As shown in Ref.~\cite{Ballesteros:2020adh}, DM particles produced in the early universe via scattering off SM particles inherit a phase space distribution different from the thermal distributions of their progenitors, which in our case can be well fitted by~\cite{Ballesteros:2020adh}
\begin{equation}
    f(q)\,\propto\,q^{-0.29}\,e^{-1.1\,q}\,,
    \label{eq:PSD}
\end{equation}
where $q$ is the DM comoving momentum defined as
\begin{equation}
  q\equiv \dfrac{p\,a(t)}{T_0} \left(\frac{g_{*s}^{\rm RH}}{g_{*s}^{0}}\right)^{1/3} \,,
\end{equation}
where $p$ is the DM momentum, $T_0$ the photon temperature at the present time, $g_{*s}^0$ and $g_{*s}^{\rm RH}$ are respectively the effective entropic degrees of freedom at present time and reheating. As the DM distribution is different from the WDM thermal distribution, the Lyman-$\alpha$ bound on the WDM mass cannot be used directly. However, it can be mapped into our scenario by using an equation-of-state matching procedure~\cite{Ballesteros:2020adh}, i.e. finding the value of $m_{\phi}$ such that
\begin{align}
w(m_\phi)=w_{\rm WDM}(m_{\rm WDM}^{\text{Ly}-\alpha})\,,
\label{eq:equalitiesw}
\end{align}
where $w\equiv \bar P/\bar \rho$ is the equation-of-state parameter defined as the ratio of the background DM pressure over background energy density which should be computed using our DM distribution of Eq.~(\ref{eq:PSD}). From the condition of Eq.~(\ref{eq:equalitiesw}), the lower bound from the Lyman-$\alpha$ analysis can be translated into our scenario as
\begin{equation}
    m_{\phi}\gtrsim 7.2~\text{keV}\, \left( \dfrac{m_{\rm WDM}^{{\rm Ly-}\alpha}}{3~\text{keV}} \right)^{4/3}  \left( \dfrac{106.75}{g_{*s}^{\rm RH}} \right)^{1/3}\,.
    \label{Eq:lyman2}
\end{equation}
Notice that taking the most or least conservative value of $m_{\rm WDM}^{{\rm Ly-}\alpha}$ in Eq.~(\ref{bound1}) represent respectively a stronger or weaker bound by a factor $\sim 2$.

\section{IV. Analysis}

\begin{figure}[ht]
\centering
\includegraphics[width=3.in]{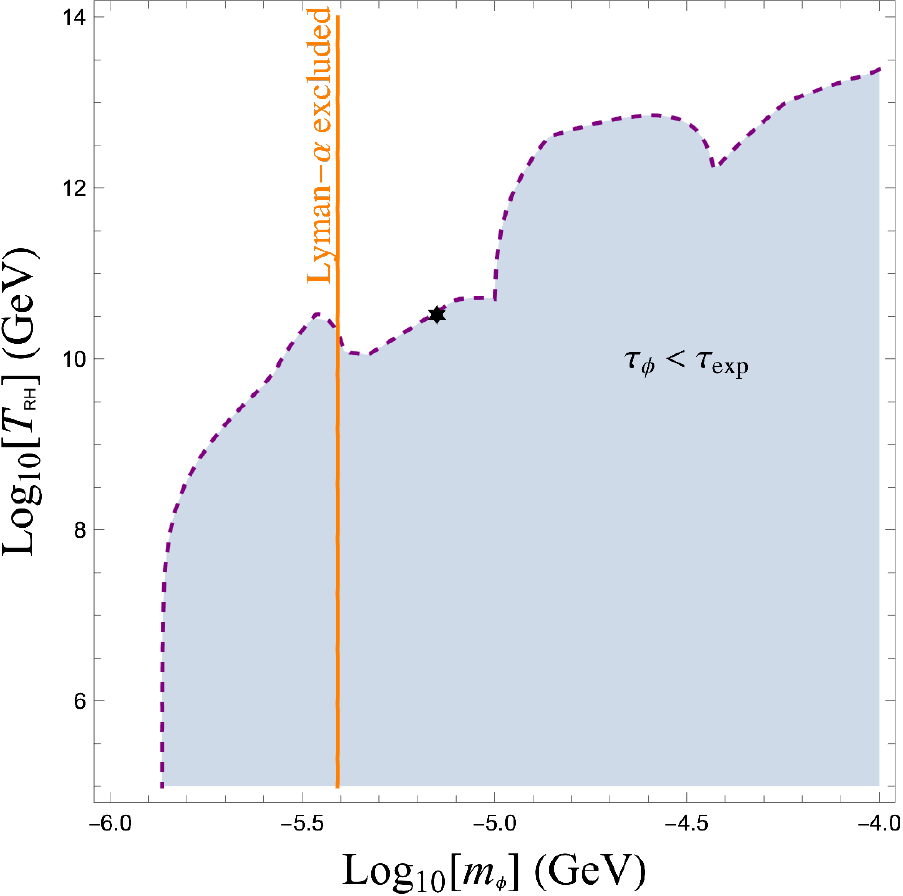}
\caption{\em \small Parameter space allowed by the combined cosmological and X-ray constraints in the plane ($m_\phi$, $\trh$). The points situated on the dashed/purple line respect the lifetime limits {\it and} $\Omega h^2\simeq0.1$, whereas the star indicates the  monochromatic signal observed by XMM-Newton \cite{Bulbul:2014sua}. The points on the left of the vertical line are excluded by the Lyman-$\alpha$ constraint (see the text for details).}
\label{Fig:omegafinal}
\end{figure}

We show in Fig.~\ref{Fig:omegafinal} the combined constraints from the dark matter lifetime and the relic abundance in the plane ($m_\phi$, $\trh$). 
The gamma-ray constraints on the lifetime of $\phi$ is extracted from different observations: XMM-Newton observations of M31~\cite{Boyarsky:2007ay} for $m_\phi \lesssim 10$ keV, NuSTAR observation of the bullet cluster~\cite{Riemer-Sorensen:2015kqa} for $m_\phi \gtrsim 10$ keV and INTEGRAL~\cite{Yuksel:2007dr} for $m_\phi \gtrsim 100$ keV\footnote{See also~\cite{DeRomeri:2020wng},~\cite{spin32}, and ~\cite{neutrino} for equivalent analysis the case of a sterile neutrino, higher-spin or Majoron DM respectively. More general cases are treated in~\cite{Chu:2012qy}}. We also show in Fig.~\ref{Fig:omegafinal} the Lyman-$\alpha$ limit obtained from Eq.~(\ref{Eq:lyman2})
taking the more conservative bound
$m_{\rm WDM}^{{\rm Ly-}\alpha}=1.9$ keV
from Eq.~(\ref{bound1}), giving $m_\phi \gtrsim 3.9$ keV.

We illustrate as a potential smoking gun signature the point (star) corresponding to a  dark matter mass $m_\phi=7.1$ keV
and a lifetime $\tau_\phi\simeq 5 \times 10^{27}$ seconds.
These values correspond to the monochromatic X-ray observation made by the satellite 
XMM-Newton interpreted as a signal of dark matter decay \cite{Bulbul:2014sua}. Notice that this benchmark point is actually in tension with the least conservative bound of Eq.~(\ref{bound1}).

The procedure to obtain this plot is straightforward.
For each dark matter mass, we extracted the {\it upper} bound on $\alpha$ from Eq.~(\ref{Eq:gammagamma}) respecting the lifetime constraints. In turn, this upper bound on $\alpha$ gives a {\it lower} bound on $\trh$, needed
to fulfill the relic abundance from Eq.~(\ref{Eq:omegabis}). The points situated on the dashed/purple line satisfy thus the lifetime limits {\it and} $\Omega_\phi h^2\simeq 0.12$, whereas the ones below the line are excluded by X-ray constraints, and the one above the line are allowed given a much longer lifetime. In other words, if a monochromatic signal is observed in the range 1 keV-1 MeV, the mass being determined by the position of the peak, and the lifetime by the height of the peak, it would be possible to deduce the reheating temperature needed to respect the cosmological limit on dark matter abundance. Hence our model is extremely predictive. Notice also that we stopped our analysis at $m_\phi \simeq 0.1$ MeV, because the reheating temperature needed to fulfill the Planck constraints is above the inflaton mass, $m_\Phi=3 \times 10^{13}$ GeV we considered  in addition to the fact that if $m_\phi>2m_e\sim$ MeV, achieving the correct relic density is incompatible with the constraints on the dark matter lifetime.

Notice that our treatment hides the dependence on $\alpha$ of the parameter space, replacing it by $\tau_\phi$. However it can be interesting to evaluate the BSM scale $\Lambda=\frac{M_P}{\alpha}$ for some benchmark points.
A point at the edge of the limit above the {\it star} in the Fig.~\ref{Fig:omegafinal} corresponds to the parameters
\bea
&&
(m_\phi, \trh, \alpha)=(7.1~\rm{keV}, 3.5\times 10^{10}~\rm{GeV}, 7400)
\nonumber
\\
&&
~\leftrightarrow (\Omega h^2,\tau_\phi)= (0.12,8.7\times 10^{27}~s.).
\nonumber
\eea
We see that the corresponding BSM scale $\Lambda \simeq 3.2\times 10^{14}~\mrm{GeV}~\gtrsim m_\Phi$, fully justifying our approach. Notice also that the more stringent  the limit from X-ray observations, the smaller the {\it upper} bound on $\alpha$, and the more consistent  our procedure, pushing the BSM scale toward the Planck scale.

\section{V. Conclusion}

In this work, we show that a dark matter candidate $\phi$, conformally coupled to the Standard Model allows for a sufficient production even for
Planck-reduced coupling. 
The high temperatures generated by the inflaton decay in the earliest
stage of reheating are sufficient to overcome the feeble coupling
whereas in the meantime, for dark matter masses below $\lesssim 1$ MeV, the lifetime is sufficiently large to respect the X-ray and $\gamma-$ray constraints
of a variety of telescope experiments. We also included the limits on $m_\phi$
from the more recent Lyman-$\alpha$ analysis, $m_\phi\gtrsim 3.9$ keV.
Our construction is very predictive as  summarized by Eq.~(\ref{Eq:lifetime}) and Fig.~\ref{Fig:omegafinal} where we exhibit the parameter space allowed and observable by the future experimental analysis of the X-ray sky. In particular, triangle loops of decoupled fermions generate decay processes of the type $\phi \rightarrow \gamma \gamma$ whose typical signature
(a monochromatic photon) is a smoking gun signal for dark matter searches. 
Such an observation, combined with the relic abundance constraint would 
determine completely the parameter space of our model. In addition to the indirect searches, new techniques of the direct detection searches for dark matter may allow us to explore the parameter space down to keV mass scales in the future~\cite{Alexander:2016aln}.

\vskip.1in
{\bf Acknowledgments:}
\noindent 
The authors want to thank especially Keith Olive, 
Emilian Dudas, Marcos Garcia, Pyungwon Ko and Dr. Biglouche for very insightful
discussions. This project has received support from the European Union’s Horizon 2020 research and innovation programme under the Marie Skodowska -Curie grant agreement No 860881-HIDDeN and the CNRS PICS MicroDark.. 
The work of MP was supported by the Spanish Agencia Estatal de Investigaci\'{o}n through the grants FPA2015-65929-P (MINECO/FEDER, UE),  PGC2018-095161-B-I00, IFT Centro de Excelencia Severo Ochoa SEV-2016-0597, and Red Consolider MultiDark FPA2017-90566-REDC. 
The work of KK was supported in part by a KIAS Individual Grant (Grant No. PG080301) at 
Korea Institute for Advanced Study.

\appendix

\section*{Appendix}

\section{A. Derivation of the Lagrangian}
\label{sec:AppLagrangian}
To make our framework clearer, we summarize here the relevant part of the Lagrangian.
The dark matter sector is constructed by introducing the conformal factor as $\tilde g_{\mu\nu}=C(\phi)g_{\mu\nu}$, where we work on the mostly-minus convention, $(+,-,-,-)$, for the metric.
To reproduce the Einstein-Hilbert action in the $g_{\mu\nu}$-frame, we define the gravity sector in the $\tilde g_{\mu\nu}$-frame by
\bea
{\cal S}_{\rm grav} = \frac{M_P^2}{2}\int \diff ^4x\sqrt{-\tilde g}C^{-1}\tilde R,
\eea
where $\tilde R$ is the Ricci scalar in the $\tilde g_{\mu\nu}$-frame.
Thus, from the relation
\begin{multline}
    C\tilde R\,=\,R+\frac{(d-1)(d-2)}{4}g^{\mu\nu}(\del_\mu\ln C)(\del_\nu\ln C)\\+(d-1)\del^2\ln C\,,
\end{multline}
for $d$-dimensional space-times (and we will take $d=4$), we obtain
\begin{equation}
{\cal S}_{\rm grav} = \frac{M_P^2}{2}\int \diff ^4x\sqrt{-g}\left[R+\frac{3}{2}g^{\mu\nu}(\del_\mu\ln C)(\del_\nu\ln C)\right].
\end{equation}
By defining the dark matter sector as
\begin{equation}
{\cal S}_{\rm DM} = \int \diff ^4x\sqrt{-\tilde g}\left[M_P^2K \tilde g^{\mu\nu}(\del_\mu C^{1/2})(\del_\nu C^{1/2})-\tilde V\right],
\end{equation}
with $K$ and $\tilde V$ being an arbitrary constant and the scalar potential in the $\tilde g_{\mu\nu}$-frame, respectively, we end up with
\begin{multline}
{\cal S}_{\rm grav}+{\cal S}_{\rm DM} = \int \diff^4x \sqrt{-g}\Big[\frac{M_P^2}{2}R+\frac{1}{2}g^{\mu\nu}(\del_\mu\phi)(\del_\nu\phi)\\-\frac{1}{2}m_\phi^2\phi^2\Big],
\end{multline}
where we have used $C(\phi)=e^{\alpha\phi/M_P}$, and $\tilde V/C^2\equiv(1/2)m_\phi^2\phi^2$.
We also assume $K=2/\alpha^2-3$, so that the kinetic term for $\phi$ takes the canonical form in the $g_{\mu\nu}$-frame. The action of the Standard Model sector is
\begin{equation}
{\cal S}_{\rm SM} = \int \diff^4x\sqrt{-\tilde g}\tilde{\cal L}_{\rm SM}\,,
\end{equation}
where the metric in $\tilde{\cal L}_{\rm SM}$ is assumed to be $\tilde g_{\mu\nu}$.
To see the interaction between $\phi$ and the Standard Model particles, it is convenient to expand $\tilde g_{\mu\nu}$ as $\tilde g_{\mu\nu}\simeq g_{\mu\nu}+\delta g_{\mu\nu}$ with $\delta g_{\mu\nu}=(\alpha\phi/M_P)g_{\mu\nu}$.
Thus, we obtain
\begin{equation}
{\cal S}_{\rm SM} \simeq \int \diff ^4 x\sqrt{-g}{\cal L}_{\rm SM} +\delta {\cal S}_{\rm SM},
\end{equation}
with
\begin{equation}
\delta {\cal S}_{\rm SM} = -\frac{\alpha}{2}\frac{\phi}{M_P}\int \diff ^4x\sqrt{-g}g^{\mu\nu}T^{\rm SM}_{\mu\nu}\,,
\end{equation}
where the SM energy-momentum tensor $T^{\rm SM}_{\mu\nu}$ is defined as
\begin{equation}
  T^{\rm SM}_{\mu\nu}   \equiv \dfrac{2}{\sqrt{-g}}\dfrac{\delta (\sqrt{-g} \mathcal{L}_\text{SM})}{\delta g^{\mu \nu}} \,, 
\end{equation}
which is explicitly written in Eqs. (\ref{eq:TSM}-\ref{eq:V(H)}).

\section{B. A possible coupling to the inflaton}
\label{sec:AppInflatonCoupling}

Assuming that the DM scalar field $\phi$ interacts with the inflaton field $\Phi$ via a renormalizable coupling (in the Einstein frame) like
   \begin{equation}
       \mathcal{V}\,=\, \kappa \phi^2 \Phi^2\,,
   \end{equation}
   which generates an effective mass for the DM during inflation
\begin{align}
    m_{\phi, \rm eff}^2\,=\,\dfrac{\partial^2 \mathcal{V}}{\partial \phi^2}\sim \kappa M_P^2  \,.
    \label{eq:DM_effectivemass}
\end{align}
\textbf{}   A contribution to the DM density is generated at the end of inflation by $4$-point processes corresponding to the time dependent dissipation rate\,\cite{GKMO2} 
   \begin{equation}
    \Gamma_{\Phi \Phi \rightarrow \phi \phi}\,=\,  \dfrac{\kappa^2 \rho_\Phi (t)}{8 \pi m_\Phi^3}\,\,
\end{equation} 
which corresponds to the following contribution to the relic density
\begin{align}
    \frac{\Omega_\phi h^2}{0.1}\simeq 
    &&\left( \dfrac{m_\phi}{10\,\rm{keV}} \right) \left( \dfrac{\kappa}{2.4 \times 10^{-4}} \right)^2 \left( \dfrac{y}{ 10^{-5}} \right) \left( \dfrac{g_s^{\rm inf}}{106.75} \right)^{3/4}\nonumber\\
    &&\times
    \left( \dfrac{m_\Phi}{3\times 10^{13}\,\rm{GeV}} \right)^{-7/2} 
    \left( \dfrac{\rho^{\rm end}_\Phi}{0.175\,m_\Phi^2 M_P^2} \right)^{1/2}
    \label{eq:DMfromkappa}
\end{align}
where $y\equiv ( 8 \pi \Gamma_\Phi/m_\Phi)^{1/2}$ is the typical dimensionless coupling of the inflaton to the SM inducing reheating. $\rho^{\rm end}_\Phi$ is the energy stored in the inflaton field at the end of inflation.

Isocurvature perturbations can be induced if the DM effective mass becomes lower than the Hubble expansion rate during inflation. According to Eq. (\ref{eq:DM_effectivemass}), this could occur only for very small values of $\kappa$. However for such small values of $\kappa$, the DM production induced by such coupling according to Eq. (\ref{eq:DMfromkappa}) would be negligible and therefore we expect both the DM background density and fluctuations to be both predominantly produced by the freeze-in mechanism. In this case, the DM fluctuations would be inherited from the SM plasma and therefore are expected to be adiabatic. A more quantitative analysis of the previous statement goes beyond this work.

\vspace{-.5cm}
\bibliographystyle{apsrev4-1}

\end{document}